\documentclass[runningheads]{llncs}

\usepackage{subcaption}
\usepackage{graphicx}
\usepackage{array}
\usepackage{multirow}
\usepackage{comment}
\usepackage[linesnumbered,ruled,vlined]{algorithm2e}

\newcommand{\norm}[1]{\| #1 \|}

\usepackage{color}

\begin{document}
\title{Self-Supervised Learning based CT Denoising using Pseudo-CT Image Pairs}
%
%

\author{Dongkyu Won \and Euijin Jung \and Sion An \and Philip Chikontwe \and
Sang Hyun Park\thanks{Corresponding author}}

\authorrunning{Won, D et al.}


\institute{Department of Robotics Engineering, DGIST, Daegu, South Korea
\email{\{won548,euijin,sion\_an,philipchicco,shpark13135\}dgist.ac.kr}}

\titlerunning{Self-supervised Pseudo-CT}


%
\maketitle              
\begin{abstract}

Recently, Self-supervised learning methods able to perform image denoising without ground truth labels have been proposed. These methods create low-quality images by adding random or Gaussian noise to images and then train a model for denoising. Ideally, it would be beneficial if one can generate high-quality CT images with only a few training samples via self-supervision. However, the performance of CT denoising is generally limited due to the complexity of CT noise. To address this problem, we propose a novel self-supervised learning-based CT denoising method. In particular, we train pre-train CT denoising and noise models that can predict CT noise from Low-dose CT (LDCT) using available LDCT and Normal-dose CT (NDCT) pairs. For a given test LDCT, we generate Pseudo-LDCT and NDCT pairs using the pre-trained denoising and noise models and then update the parameters of the denoising model using these pairs to remove noise in the test LDCT. To make realistic Pseudo LDCT, we train multiple noise models from individual images and generate the noise using the ensemble of noise models. We evaluate our method on the 2016 AAPM Low-Dose CT Grand Challenge dataset. The proposed ensemble noise model can generate realistic CT noise, and thus our method significantly improves the denoising performance existing denoising models trained by supervised- and self-supervised learning.

\end{abstract}
%
%


\section{Introduction}

Computed Tomography (CT) is a widely used medical imaging modality to visualize patient anatomy using X-rays. However, X-rays are known to be harmful to the human body as radiation may cause genetic and cancerous diseases. To address this, Low-dose CT (LDCT) images are often taken by lowering the X-ray exposure to reduce the dose yet inevitably inducing higher noise with lower quality. Though this mitigates the potential side-effects to patients, LDCT images are insufficient for accurate medical diagnosis compared to high quality Normal-dose CT (NDCT) images. Thus, robust denoising techniques are required to enhance the quality of LDCT.

In literature, several studies have attempted to address CT denoising using filtering  \cite{chen2012thoracic,ma2011low,li2014adaptive,feruglio2010block,kang2013image}, dictionary learning \cite{chen2013improving,zhang2015projection}, and deep learning methods \cite{chen2017low,yang2017ct,chen2017lowresidual,won2020low}, respectively. Broadly speaking, existing methods try to maximize some prior given well defined noise model measurements and also exploit correlations among pixels to regulate the denoised image. For data-driven approaches, human experts are often required to mathematically model pixel correlations based analysis of clean images; limiting applicability when data is scarce. Thus, deep learning methods~\cite{chen2017low,yang2017ct,chen2017lowresidual,won2020low,nishio2017convolutional,wolterink2017generative,zhong2020image} are attractive given the recent impressive results for CT denoising, especially the supervised learning methods.

Despite the continued mainstream adoption of deep learning based techniques for medical image analysis, performance is still hindered by the availability of large and rich amounts of annotated data. For example, achieving high quality denoising requires several paired LDCT and NDCT images for models to learn an optimal mapping. Indeed, several techniques\cite{krull2019noise2void,lehtinen2018noise2noise} have opted to train a denoiser without ground-truth labels for natural images where assumptions about noise are modeled based on surrounding pixel values including masking strategies to enforce the model to learn data complexity. However, such assumptions may not hold for medical images. 

In this paper, we propose a novel self-supervised learning framework for CT denoising using Pseudo-CT noise images. We aim to directly address the limited availability of LDCT/NDCT images by defining a self-supervised task on a pair of LDCT and Pseudo noise CT generated images that can model different types of unseen noise during training. Specifically, we first pre-train a CT denoiser that predicts NDCT from LDCT, and also pre-train a LDCT noise model tasked with generating a noise difference map i.e. NDCT-LDCT, which we term Pseudo-CT for simplicity. Second, we train a new denoiser initialized with weights of the pre-trained denoiser by iteratively optimizing a self-supervised task between its output and the Pseudo-CT. Through learning to generate different types of unseen noise by ensembling across noise predictions for self-supervision, our approach produces efficient and high-quality denoising models. The main contributions of this work are as follows: (1) we propose a novel self-supervised learning scheme for CT denoising, (2) we show that our strategy can create realistic Pseudo-LDCT by leveraging diverse noise generated from ensemble noise models and, (3) Empirical results confirm that our method improves the performance of several existing methods.

\section{Related work}
In this section, we discuss recent advances for CT denoising in two aspects; supervised LDCT denoising\cite{chen2017low,yang2017ct,chen2017lowresidual,won2020low,nishio2017convolutional,wolterink2017generative,zhong2020image} and self-supervised denoising methods \cite{lehtinen2018noise2noise,batson2019noise2self,krull2019noise2void,lee2020restore}.

\subsubsection{LDCT Denoising} 
Here, we first highlight conventional CT denoising techniques such as filtering \cite{chen2012thoracic,ma2011low,li2014adaptive,feruglio2010block,kang2013image} and dictionary based learning \cite{chen2013improving,zhang2015projection}. These techniques generally employ informative filters or dictionary patches to achieve denoising. Though they show impressive results, performance is largely limited by difficulty in modeling complex CT noise distributions.

Recently, several deep learning-based methods\cite{chen2017lowresidual,yang2017ct,won2020low} have emerged for CT denoising. For instance, Chen et al.\cite{chen2017low} proposed a simple shallow convolutional neural network(CNN) with three layers to remove noise. Following, Yang et al.\cite{yang2017ct} proposed a deeper model with perceptual loss to better preserve structural information. A more recent work by Won et al. \cite{won2020low} adopted intra- and inter-convolutions to extract features in both low and high-frequency components of CT images. As for generative approaches, Nishio et al. \cite{nishio2017convolutional} proposed a CNN-based auto-encoder using NDCT with noise as input and predicts denoised NDCT. Similarly, Wolterinkt et al.\cite{wolterink2017generative} proposed a generative adversarial network(GAN) based approach for denoising. The generator generates CT noise and then subtracts from LDCT images to generate NDCT images, with a discriminator used to classify the generated NDCT as real or fake. However, these methods rely on the availability of several LDCT and NDCT paired samples to be successful. 

\subsubsection{Self-Supervised Denoising}
To address the shortcomings of supervised methods, self-supervised techniques are very attractive for CT denoising. These methods assume paired data is unavailable for training and define self-supervised tasks to achieve denoising performance on par with supervised methods. Lehtinen et al.\cite{lehtinen2018noise2noise} proposed Noise2Noise; a novel strategy to predict a clean image only using noisy image pairs. Furthermore, Batson et al.\cite{DBLP:conf/icml/BatsonR19} (Noise2Self) and Krull et al.\cite{krull2019noise2void} (Noise2Void) both proposed masking strategies for denoising using only noisy images. These works specifically addressed the blind-spot problem; a common phenomena where a networks produces degenerative predictions by learning the identity in limited pixel receptive fields. By hiding or replacing specific pixels in the images, models are able to remove pixel-wise independent noise showing improved performance. Taking a step forward with the methods above, Lee et al. \cite{lee2020restore} proposed a self-supervised denoising strategy by updating denoiser with pseudo image pairs. They stated self-similar images help improving the performance and they generated pseudo image pairs consist of pseudo clean images from pre-trained denoiser and pseudo noisy images adding random gaussian noise to the clean image. This seems reasonable for natural images, yet is unclear if such approaches can be directly applied to the medical domain.

Nevertheless, Hasan et al.\cite{hasan2020hybrid} made early attempts with Noise2Noise\cite{lehtinen2018noise2noise} for CT denoising. Here, several generators networks called hybrid-collaborative generators were proposed to predict NDCT from many LDCT pairs. Hendriksen et al.\cite{hendriksen2020noise2inverse} employed Noise2Self for sinogram reconstruction by dividing sinograms into several groups and later predict a sub-sinogram from the sub-sinograms. Notably, the borrowed techniques are domain specific solutions for denoising in natural images. Consequently, this may degrade denoising performance in CT and they cannot accurately reflect noise properties observed in CT. We address this by defining a new self-supervised learning scheme with iterative updates for model training. 
\section{Method}

\begin{figure}[t]
\centering
\includegraphics[width=1.0\textwidth]{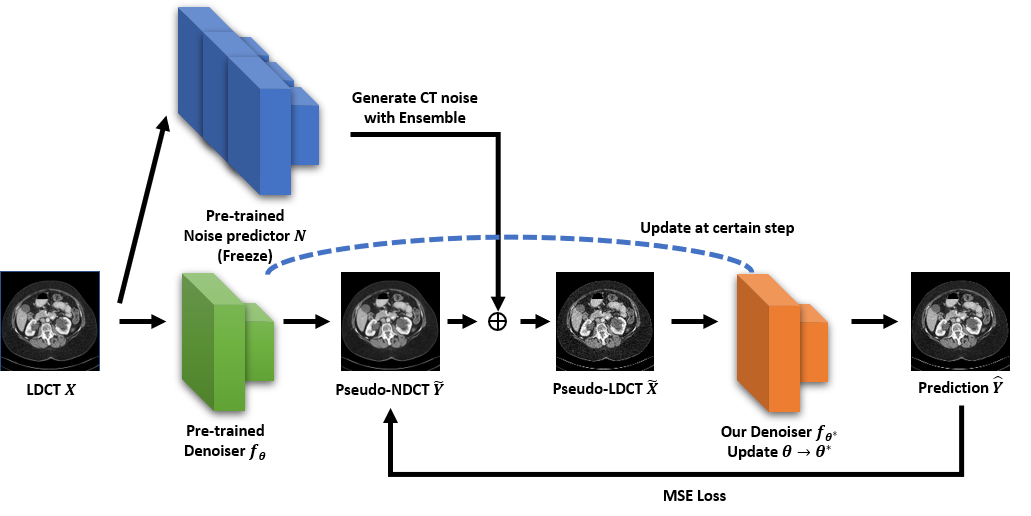}
\caption{Proposed self-supervised LDCT denoising method.} \label{fig::method}
\end{figure}

Let us assume we already have a pre-trained LDCT denoiser $f_\theta$ trained with LDCT ${X}$ and NDCT $Y$ images i.e. $f_\theta$ can be trained with any learning method, supervised or self-supervised. Our goal is to denoise an unseen LDCT test image via self-supervised fine-tuning of $f_\theta$ with Pseudo noise CTs. To achieve this, we also pre-train a LDCT noise model $N$ to generate Pseudo noisy LDCT $\tilde{X}$ using the difference map between LDCT and NDCT as supervison, with $f_\theta$ producing $\tilde{Y}$ i.e. Pseudo-NDCT. Following, to train an improved CT denoiser $f(\theta^\star)$ initialized the weights of $f_\theta$, we simultaneously update $f_\theta$ and $f(\theta^\star)$ at different time steps using a self-supervised loss between the Pseudo-CT images and the predictions i.e. $\hat{Y} = f(\theta^\star)$. Our proposed method is shown in Figure 1 with specific details expanded below.

\noindent{\textbf{{Pre-training Noise Models}}.
The proposed method requires a pre-trained LDCT noise model $N$ to predict CT noise from $X$. Here, $N$ is trained to predict noise map $Z$, which is the difference image between $X$ and $Y$. In particular, we constructed pairs $\{X,Z\}$ to train $N$ with each noise map obtained by subtracting $X$ from $Y$. For this work, we train several $N$ networks equal to the number of CT subjects in the training dataset used by $f_\theta$ i.e. we used 3 LDCT noise models. For all subsequent procedures, these models are fixed without parameter update.

\noindent{\textbf{{Pseudo-CT Image Generation}}.
To generate pseudo images for supervision, both $f_\theta$ and $N$ are employed. Pseudo-CT images consist of a pair of $\tilde{X}$ and $\tilde{Y}$ and are generated as follows,

\begin{equation}
\tilde{Y}=f_\theta(X),
\end{equation}
\begin{equation}
Z_m=N_m({X)},
\end{equation}
\begin{equation}
\tilde{X}=\tilde{Y}+Ensemble(Z),
\end{equation}
where $m$ is the index of $N$, with $f_\theta$ used to generate $\tilde{Y}$. As mentioned, $f_\theta$ is a pre-trained model tasked to predict $X$ as $Y$. Once $X$ is fed to $f_\theta$, we obtain a clean version of $X$ and denote this as Pseudo-NDCT $\tilde{Y}$. Also, the ensemble of $N$ predictions is applied to generate $\tilde{X}$ by feeding $X$ into $N_m$ to predict noise $Z_m$. Specifically, to generate $Z$ by ensembling; we randomly select pixels among $Z_m$ for all required pixels. Repeating this process several times enables us to create diverse noise maps per input facilitating stable learning. Thus, by adding $Z$ to $\tilde{Y}$ we can obtain the final Pseudo-LDCT image $\tilde{X}$. During training, only $\tilde{X}$ and $\tilde{Y}$ will be used for parameter updates.

\SetKwInput{KwHyper}{Hyperparameter}
\SetKwInput{KwInput}{Input}
\SetKwInput{KwOutput}{Output}
\SetKwInput{KwNet}{Pre-trained model}
\begin{algorithm}[t]
\DontPrintSemicolon
  \KwInput{Low-Dose CT image X}
  \KwOutput{Denoised Image $\hat{Y}$}
  \KwNet{denoising model $f_\theta$, noise models $\big\{N_1, \cdot\cdot\cdot, N_m\big\}$ where $m$ is the number of training set}
  \KwHyper{learning rate $\alpha$, update period $C$}
   \For{$i \gets$ 1 to m}
   {
        Generate noise $Z_i=N_i(X)$\\
   }
   $count = 0$\\
   $\theta^\star \gets \theta$\\
   \While{not done}
   {
        Generate Pseudo-NDCT image $\tilde{Y} = f_\theta(X)$\\
        Generate Pseudo-LDCT image $\tilde{X}=\tilde{Y}+Ensemble(Z)$\\
        Parameter update $\theta^\star \gets \theta^\star + \alpha\bigtriangledown_{\theta^\star}\mathcal{L}(\tilde{Y}, f_{\theta^\star}(\tilde{X}))$\\
        $count \gets count + 1$\\
        \If{$count\equiv$ 0 $\pmod{C}$}
        {
             $\theta \gets \theta^\star$\\
        }
   }
  Return $\hat{Y} = f_{\theta}(X)$
\caption{Parameter update procedure}
\end{algorithm}

\noindent{\textbf{Pseudo-CT Self-Supervision}}.
Given the generated Pseudo-CT images, improved denoising model $f(\theta^\star)$ can be optimized via self-supervision. First, $f(\theta^\star)$ is initialized with the parameters $f_\theta$, with $f_\theta$ and $N$ initially frozen when generating $\tilde{X}$ and $\tilde{Y}$ at each training step. Formally, $f(\theta^\star)$ is optimized with the loss:
\begin{equation}
L(\tilde{Y}, \hat{Y})= \frac{1}{K} \sum \limits_{k=1}^{K} \norm{ f_{\theta^\star}(\tilde{X}) - \tilde{Y} }_2^2.
\end{equation}
where $\hat{Y}$ is prediction from $f(\theta^\star)$ and $K$ is the mini-batch size per step. In order to generate and provide high-quality Pseudo-LDCT images to $f(\theta^\star)$, we update $\theta$ with $\theta^\star$ after a fixed number of steps. This allows $f_\theta$ to provide continuously improved Pseudo-CT images to $f(\theta^\star)$ and eventually improves CT denoising performance. Algorithm 1 provides the details of our method.

\noindent{\textbf{{Implementation Details}}.
For this work, we used REDCNN\cite{chen2017lowresidual} as the pre-trained LDCT denoising model and our final CT denoiser, with U-Net\cite{Ronneberger2015} as the LDCT noise model. All methods employed CT patches of $64\times64$ in size. Data augmentation was used in all training and update steps. For data augmentation, random rescaling between [0.5, 2], and random flipping [Horizontal, Vertical] was used. An L1 loss was used to train the CT denoising and LDCT noise models, with Adam optimizer and a learning rate (LR) of 0.0001. A ReduceLROnPlateau scheduler was employed to adjust the LR by 50\% of the initial value whenever the loss did not improve during training. All experiments were implemented on a Intel Xeon Gold 6132, 192GB RAM, NVIDIA TITAN Xp workstation with the PyTorch deep learning framework.

\begin{table}[t]
\centering
\caption{Quantitative results on supervised and self-supervised method and various Pseudo-CT image generation.} \label{tab1}
\begin{tabular}{l|c|c}
\hline
Method 	 							& PSNR(std) & SSIM(std) \\
\hline\hline
NDCT   				 				& ${46.21 \pm 1.37}$&${0.9773 \pm 0.0072}$ \\ 

\hline\hline
N2V \cite{krull2019noise2void}      & ${46.46 \pm1.23}$&${0.9790 \pm 0.0054}$ \\
N2V+Ours w/o $\theta$ Update        & ${49.22 \pm 0.82}$&${0.9893 \pm 0.0027}$ \\
\textbf{N2V+Ours}   & \boldmath{${49.57\pm0.82}$}&\boldmath{${0.9905\pm0.0019}$} \\ 
\hline\hline
N2N \cite{lehtinen2018noise2noise}  & ${46.16\pm0.93}$&${0.9769 \pm 0.0039}$ \\
N2N+Ours w/o $\theta$ Update        & ${46.69\pm1.14}$&${0.9794 \pm 0.0055}$ \\
\textbf{N2N+Ours}   				& \boldmath{${49.85\pm0.86}$}&\boldmath{${0.9910\pm0.0018}$} \\ 
\hline\hline
N2C \cite{chen2017lowresidual}      & ${50.30\pm0.93}$&${0.9920 \pm 0.0018}$ \\
N2C+Ours w/o $\theta$ Update	    & ${50.47\pm0.98}$&${0.9922 \pm 0.0018 }$ \\
\textbf{N2C+Ours}   				& \boldmath{${50.50\pm0.98}$}&\boldmath{${0.9922\pm0.0018}$} \\ 
\hline\hline
N2C+Hist                        	& ${48.65\pm1.42}$&${0.9872 \pm 0.0045}$ \\
N2C+Gaussian                        & ${49.94\pm0.87}$&${0.9911 \pm 0.0019}$ \\
N2C+Model+Hist                      & ${50.36\pm0.95}$&${0.9920 \pm 0.0018}$ \\
\hline
\end{tabular}
\end{table}

\begin{figure}[t]
\makebox[\linewidth][c]{%
\begin{subfigure}[b]{.25\textwidth}
\centering
\includegraphics[width=0.95\textwidth]{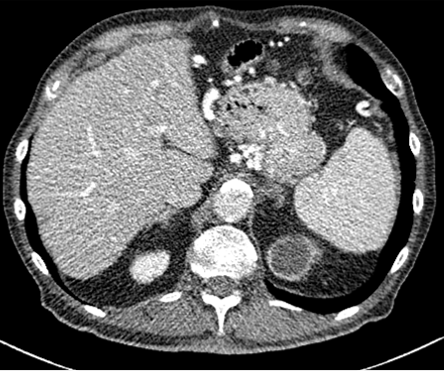}
\end{subfigure}%
\begin{subfigure}[b]{.25\textwidth}
\centering
\includegraphics[width=0.95\textwidth]{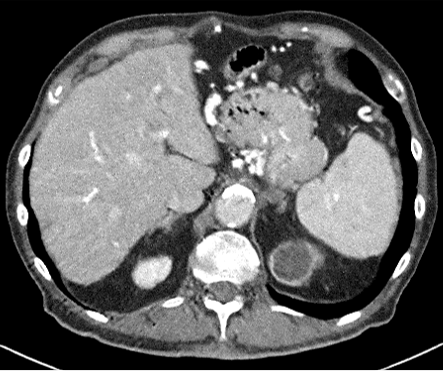}
\end{subfigure}%
\begin{subfigure}[b]{.25\textwidth}
\centering
\includegraphics[width=0.95\textwidth]{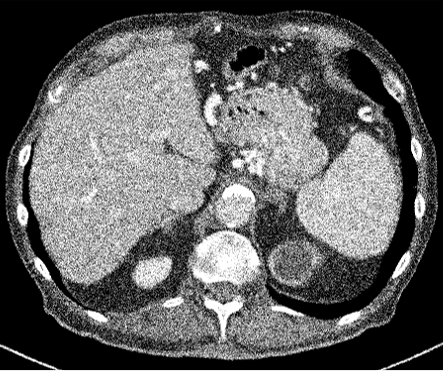}
\end{subfigure}%
\begin{subfigure}[b]{.25\textwidth}
\centering
\includegraphics[width=0.95\textwidth]{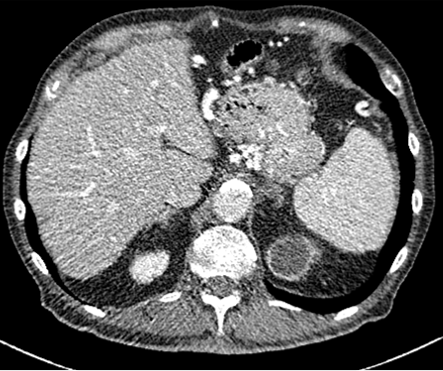}
\end{subfigure}%
}\\
\makebox[\linewidth][c]{%
\begin{subfigure}[b]{.25\textwidth}
\centering
\includegraphics[width=0.95\textwidth]{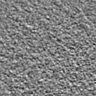}
\caption{}
\end{subfigure}%
\begin{subfigure}[b]{.25\textwidth}
\centering
\includegraphics[width=0.95\textwidth]{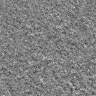}
\caption{}
\end{subfigure}%
\begin{subfigure}[b]{.25\textwidth}
\centering
\includegraphics[width=0.95\textwidth]{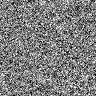}
\caption{}
\end{subfigure}%
\begin{subfigure}[b]{.25\textwidth}
\centering
\includegraphics[width=0.95\textwidth]{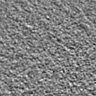}
\caption{}
\end{subfigure}%
}\\
\caption{Comparison of generated Pseudo-LDCT images and their noise. (a) LDCT, (b) Random noise histogram, (c) Gaussian noise, (d) Ours. The right-bottom patch indicates the noise lying on each image. The noise quality of (b) and (c) shows that random noise cannot represent (a). In contrast, (d) shows almost similar to (a)}
\end{figure}

\section{Experimental Results}

\subsubsection{Dataset and Experimental Settings} 
For evaluation, the 2016 AAPM Low-Dose CT Grand Challenge dataset\cite{mccollough2016tu} was used and split into 3 train and 7 test, respectively. It consists of abdominal LDCT and NDCT images obtained from 10 patients with image size $512\times512$. The voxel space of CT images is $0.5mm\times0.5mm$ with $3mm$ slice thickness. To demonstrate that our method works with any existing learning method, we applied our method to N2C (Noise2Clean), N2N (Noise2Noise)\cite{lehtinen2018noise2noise}, and N2V (Noise2Void)\cite{krull2019noise2void}. N2C is a supervised learning method that uses all available labeled paired data. N2N and N2V are self-supervised learning methods which use noisy pair images and masking schemes. Also, to demonstrate the effectiveness of our noise model, we compared LDCT denoising performance against existing noise generation techniques i.e. Random Noise histogram (Hist), Gaussian noise (Gaussian), and Single noise model$+$Noise histogram (Model+Hist). Hist samples noise from the difference map based on the histogram between LDCT and NDCT images, whereas gaussian samples the noise from a Gaussian distribution with zero mean and 0.02 standard deviation. Model+Hist is the combination of a single pre-trained LDCT noise model (using all subjects in the training dataset) and Hist. For evaluation, Peak Signal-to-Noise Ratio (PSNR) and Structural Similarity (SSIM) are reported.

\begin{figure}[t]
\makebox[\linewidth][c]{%
\begin{subfigure}[b]{.25\textwidth}
\centering
\includegraphics[width=0.95\textwidth]{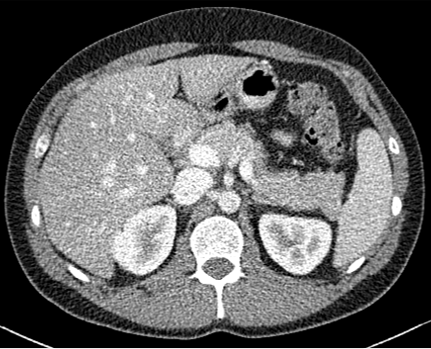}
\caption{LDCT}
\end{subfigure}%
\begin{subfigure}[b]{.25\textwidth}
\centering
\includegraphics[width=0.95\textwidth]{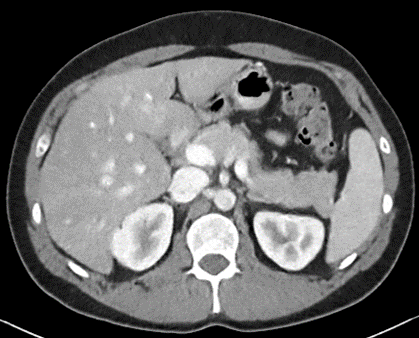}
\caption{N2C}
\end{subfigure}%
\begin{subfigure}[b]{.25\textwidth}
\centering
\includegraphics[width=0.95\textwidth]{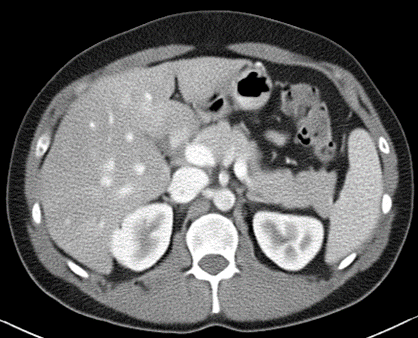}
\caption{N2N}
\end{subfigure}%
\begin{subfigure}[b]{.25\textwidth}
\centering
\includegraphics[width=0.95\textwidth]{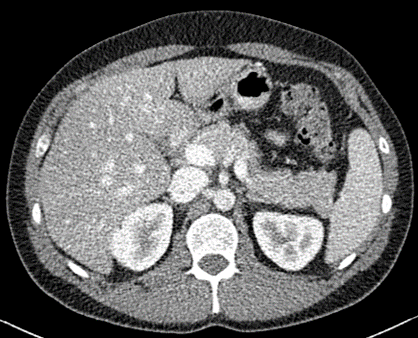}
\caption{N2V}
\end{subfigure}%
}\\

\makebox[\linewidth][c]{%
\begin{subfigure}[b]{.25\textwidth}
\centering
\includegraphics[width=0.95\textwidth]{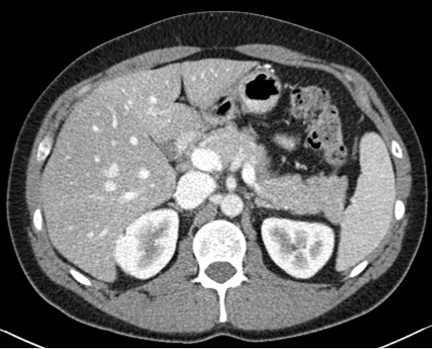}
\caption{NDCT}
\end{subfigure}%
\begin{subfigure}[b]{.25\textwidth}
\centering
\includegraphics[width=0.95\textwidth]{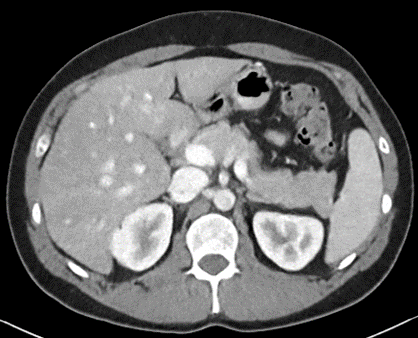}
\caption{N2C+Ours}
\end{subfigure}%
\begin{subfigure}[b]{.25\textwidth}
\centering
\includegraphics[width=0.95\textwidth]{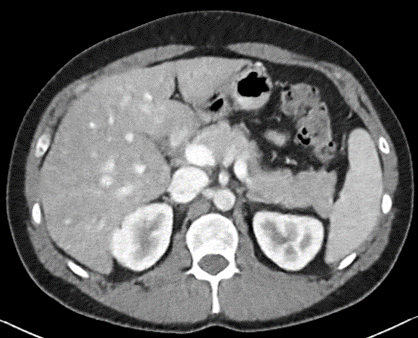}
\caption{N2N+Ours}
\end{subfigure}%
\begin{subfigure}[b]{.25\textwidth}
\centering
\includegraphics[width=0.95\textwidth]{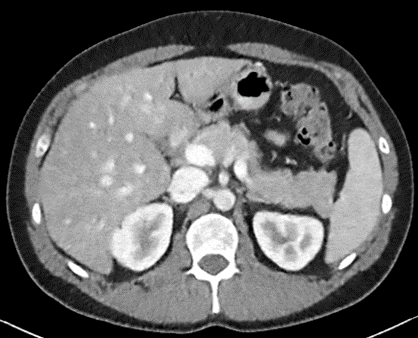}
\caption{N2V+Ours}
\end{subfigure}%
}\\
\caption{Comparison of existing methods and Ours. (a) LDCT, (b) N2C, (c) N2N, (d) N2V, (e) NDCT, (f) N2C+Ours, (g) N2N+Ours, (h) N2V+Ours.}
\end{figure}

\subsubsection{Quantitative Results} 
Table. 1 shows the average PSNR and SSIM scores for supervised and self-supervised methods. Among the compared methods, N2C showed the highest LDCT denoising performance without the proposed techniques; with more significant improvements for all methods when the our methods were applied. Though N2N and N2V report higher scores than NDCT, with similar trends on natural image denoising, they were not very effective when applied to CT. We believe training by merely adding random noise (e.g., Gaussian), often  very different from noise observed in CT is not useful for denoising, leading to poor results. Indeed, our method improved every learning method's performance showing that the technique is model agnostic.

\subsubsection{Effect of Noise Model} 
For parameter updates, it is crucial to generate high-quality Pseudo-LDCT images i.e. images similar to an actual LDCT image. If high-quality Pseudo-LDCT images are provided to the model for training, CT denoising performance can be improved. In contrast, using low-quality Pseudo-CT images may adversely affects the model learning, and consequently lead to decreased denoising performance. Table. 1 also shows the average PSNR and SSIM in different Pseudo-CT image generation settings. Here, both Hist and Gaussian reported lower performance compared to N2C. Based on our observations, this serves to show that random noise employed in natural images is not useful for CT denoising. In contrast, when our noise model was combined with Hist i.e. Model+Hist, performance gains over N2C were noted. This implies that our noise model can generate reasonable noise similar to the actual CT noise, and also improves performance. Furthermore, our ensemble noise models without parameter update show improved  results over Model+Hist without the need for additional random noise. 

In Figures 2 and 3, we show comparison results of Pseudo-LDCT images and their generated noise using various methods, as well as the predictions of our method for each. In Fig. 2, Pseudo-LDCT images and their noise with Hist and Gaussian highlight a huge discrepancy between the actual LDCT image and its noise. In the case of random noise that is shown to be independently distributed across the entire image, it is often useful for natural image denoising. Moreover, this types of noise does not accurately reflect the nature of CT images. Through the proposed method, we show it is possible to generate Pseudo-CT images that preserve overall CT image characteristics with high quality (Fig. 3).

\section{Conclusions}
In this paper, we proposed a novel self-supervised CT denoising method using Pseudo-CT images. We showed by using a noise model and generated Pseudo-CT images to define a self-supervised learning scheme, we can produce robust denoising model without relying on a large collection of labeled images. Also, we provide concrete evidence via extensive experimentation that our approach is model-agnostic and can improve CT denoising performance for existing approaches.

\bibliographystyle{splncs04}
\bibliography{bibliography}
\end{document}